# Multiple-Quantum Spin Dynamics of Entanglement

S. I. Doronin

*Institute of Problems of Chemical Physics, Russian Academy of Sciences, Chernogolovka, 142432, Moscow Region, Russia*

Dynamics of entanglement is investigated on the basis of exactly solvable models of multiple-quantum (MQ) NMR spin dynamics. It is shown that the time evolution of MQ coherences of systems of coupled nuclear spins in solids is directly connected with dynamics of the quantum entanglement. We studied analytically dynamics of entangled states for two- and three-spin systems coupled by the dipole-dipole interaction. In this case dynamics of the quantum entanglement is uniquely determined by the time evolution of MQ coherences of the second order. The real part of the density matrix describing MQ dynamics in solids is responsible for MQ coherences of the zeroth order while its imaginary part is responsible for the second order. Thus, one can conclude that dynamics of the entanglement is connected with transitions from the real part of the density matrix to the imaginary one and vice versa. A pure state which generalizes the *GHZ* and *W* states is found. Different measures of the entanglement of this state are analyzed for three-partite systems.

## I. INTRODUCTION

The solid state NMR is one of the most promising candidates for a realization of a quantum computer and quantum-information processing (QIP). Several such schemes were suggested [1-8]. New possibilities in experimental realizations of quantum algorithms and theoretical investigations of QIP are connected with multiple- quantum (MQ) NMR. Contrary to the usual NMR, MQ NMR spectroscopy uses not only adjacent but all possible transitions between Zeeman levels of the system of interacting spins in an external magnetic field. This allows to gain more detailed information on dynamical processes in solids. Further progress in the understanding of multiple-spin MQ dynamics was achieved with experimental investigations of MQ NMR dynamics of quasi-one-dimensional distributions of clusters of uniformly spaced proton spins in hydroxyapatite, $Ca_5OH(PO_4)_3$, and related fluorine-containing apatites [9,10]. These materials represent a possible base for QIP devices [6,7]. Thus, investigations of the quantum entanglement with MQ NMR methods are of current interest.

The numerical methods [11-13] and exactly solvable models [14-17] have been applied to the theoretical analysis of MQ dynamics. In this paper, we investigate dynamics of the entangled states on the basis of MQ NMR methods. Obtained results show that a time evolution of MQ coherences of nuclear spins- 1/2 coupled by the dipole-dipole interactions (DDI) is directly connected with dynamics of quantum entanglement. The exact solutions are presented for two and three spin-1/2 nuclei. For these systems the periodic dynamics of quantum entanglement is uniquely determined by the time evolution of the intensity of MQ coherences of the second order. We suppose that the intensity of MQ coherence of the second order can be considered as a measure of entanglement. In particular, it is significant that the entanglement and intensity of MQ coherence of the second order are equal to zero at the moment of time $\tau=0$. At the same time, the intensity of MQ coherence of the zeroth order is maximal at $\tau=0$. In the course of the time evolution the exchange between MQ coherences of the zeroth- and second orders takes place while their sum is a constant of motion [18]. When the intensity of MQ coherence of the zeroth order is decreasing and is getting minimal the entanglement achieves its maximum value as so does the intensity of MQ coherence of the second order. After that the entanglement decreases and so on. It was shown [11] that the real part of the density matrix describing MQ dynamics in solids is responsible for MQ coherences of the zeroth order while its imaginary part is responsible for the second order. It is possible to conclude that dynamics of the entanglement is connected with the transitions between the real and imaginary parts of the density matrix.

## II. MQ DYNAMICS

MQ spin dynamics in solids is described by the Liouville - von Neumann equation

$$i\frac{d\rho}{d\tau} = [H, \rho(\tau)], \tag{1}$$

where the nonsecular average dipolar Hamiltonian $H$ is given by [14]

$$H = -\frac{1}{2}\sum_{j<k} D_{jk}(I_j^+ I_k^+ + I_j^- I_k^-) \tag{2}$$

and $I_j^\pm$ are the raising and lowering spin angular momentum operators of spin $j$. The dipolar coupling constant $D_{jk}$ between spins $j$ and $k$ is given by

$$D_{jk} = \frac{\gamma^2 \hbar}{2r_{jk}^3}(1 - 3\cos^2\theta_{jk}), \tag{3}$$

where $r_{jk}$ is the distance between spins $j$ and $k$, $\theta_{jk}$ the angle between the internuclear vector $\vec{r}_{jk}$ and the external magnetic field $\vec{H}_0$, and $\gamma$ the gyromagnetic ratio. The angle $\theta_{jk}$ is the same for all spin pairs. In linear chains we assume $\theta_{jk}=0$.

If the spin system is in thermal equilibrium with the lattice at $\tau=0$, then the solution of Eq. (1) in the high temperature approximation is

$$\rho(\tau) = e^{-iH\tau}\rho(0)e^{iH\tau}. \tag{4}$$

Here $\rho(0)$ is the initial density matrix:

$$\rho(0) = \sum_j^N I_{zj}, \tag{5}$$

where $I_{zj}$ is the projection of spin $j$ angular momentum on the direction $z$ of the external field $\vec{H}_0$. According to [14] the density matrix $\rho(\tau)$ can be expanded as

$$\rho(\tau) = \sum_n \rho_n(\tau), \tag{6}$$

where $\rho_n(\tau)$ is the contribution to $\rho(\tau)$ due to MQ coherences of order $n$. If only the nearest neighbors interact the profile of MQ coherences consists of 0- and 2-quantum coherences [15,13]. At the same time, MQ NMR spectra consist of MQ coherences of arbitrary even orders ($n=0, \pm 2, \pm 4, \pm 6,\ldots,$ $|n|\leq N$, where $N$– number of spins) as the DDI of all spins are taken into account. While intensities of MQ coherences of order $n=4k$ ($0\leq 4k\leq N$) are determined by the real part of the density matrix $\rho(\tau)$ only, intensities of MQ coherences of the order $n=4k+2$ ($2\leq 4k+2\leq N$) are determined by the imaginary part of $\rho(\tau)$ [11].

Special experimental methods [19-21] are used to separate the signals from the MQ coherences of various orders. In the MQ NMR experiments the multiple-quantum coherences are transformed into the longitudinal magnetization. As the result, the intensities of the MQ coherences $J_n(\tau)$ can be measured. They can be expressed as [14]

$$J_n(\tau) = \mathrm{Tr}\rho_n(\tau)\rho_{-n}(\tau). \tag{7}$$

It is worth to notice that $\rho_{-n}(\tau) = \rho_n^+(\tau)$.

The sum of the intensities of the multiple-quantum coherences does not depend on time [18]. Normalizing this sum to unity, we have

$$J_0(\tau) + 2\sum_{n=2}^N J_n(\tau) = 1. \tag{8}$$

## III. DYNAMICS OF ENTANGLEMENT BETWEEN TWO SPIN NUCLEI

A pair of spins has four possible basis product states: $|00\rangle$, $|01\rangle$, $|10\rangle$, $|11\rangle$. In this basis the initial density matrix from Eq. (5) is

$$\rho(0) = \begin{pmatrix} 1 & 0 & 0 & 0 \\ 0 & 0 & 0 & 0 \\ 0 & 0 & 0 & 0 \\ 0 & 0 & 0 & -1 \end{pmatrix}. \tag{9}$$

The density matrix $\rho(\tau)$ can be obtained from the analytical solution Eq. (4) [15, 22].

$$\rho(\tau) = \rho_0(\tau) + \rho_{\pm 2}(\tau) = \begin{pmatrix} \cos\varphi(\tau) & 0 & 0 & i\sin\varphi(\tau) \\ 0 & 0 & 0 & 0 \\ 0 & 0 & 0 & 0 \\ -i\sin\varphi(\tau) & 0 & 0 & -\cos\varphi(\tau) \end{pmatrix}, \tag{10}$$

where $\varphi(\tau) = D_{12}\tau$ and $\rho_0(\tau) = \text{Re}[\rho(\tau)]$, $\rho_{\pm 2}(\tau) = \rho_{+2}(\tau) + \rho_{-2}(\tau) = \text{Im}[\rho(\tau)]$.

The normalized intensity of MQ coherences can be calculated from Eqs. (7), (10):

$$J_0(\tau) = \cos^2\varphi(\tau), \tag{11}$$
$$J_2(\tau) = J_{+2}(\tau) + J_{-2}(\tau) = \sin^2\varphi(\tau). \tag{12}$$

The density matrix $\rho(\tau)$ of Eq. (10) can be expressed through the vector of the state

$$|\Psi\rangle = e^{i\frac{\pi}{4}}\cos[\varphi(\tau)/2]\,|00\rangle + e^{-i\frac{\pi}{4}}\sin[\varphi(\tau)/2]\,|11\rangle. \tag{13}$$

The projector is

$$\sigma(\tau) = |\Psi\rangle\langle\Psi| = \frac{1}{2}\begin{pmatrix} 1+\cos\varphi(\tau) & 0 & 0 & i\sin\varphi(\tau) \\ 0 & 0 & 0 & 0 \\ 0 & 0 & 0 & 0 \\ -i\sin\varphi(\tau) & 0 & 0 & 1-\cos\varphi(\tau) \end{pmatrix}. \tag{14}$$

Comparing (10) and (14), we have

$$\rho(\tau) = 2\sigma(\tau) - E', \tag{15}$$

where $E' = \begin{pmatrix} 1 & 0 & 0 & 0 \\ 0 & 0 & 0 & 0 \\ 0 & 0 & 0 & 0 \\ 0 & 0 & 0 & 1 \end{pmatrix}$ is independent of time.

Thus, MQ dynamics of the two-spin systems can be described by the state vector (13).
Let us consider dynamics of the entanglement of the bipartite system in the state of Eq. (13).

The measure of the entanglement for a pure bipartite state can be defined as the von Neumann entropy of any of its two parts [23].

$$E = -\text{Tr}\,\sigma_A \log_2 \sigma_A = -\text{Tr}\,\sigma_B \log_2 \sigma_B, \tag{16}$$

where $\sigma_{A(B)} = \text{Tr}_{B(A)}\,\sigma$ is the reduced density matrix.

If the quantum state is given by

$$|\Psi\rangle_{AB} = a\,|0\rangle_A \otimes |0\rangle_B + b\,|1\rangle_A \otimes |1\rangle_B, \tag{17}$$

then the operator $\sigma_A$ can be written in the form (see Chapter 2.3.1 in Ref. [24])

$$\sigma_A = |a|^2\,|0\rangle_{A\,A}\langle 0| + |b|^2\,|1\rangle_{A\,A}\langle 1|. \tag{18}$$

In our case $|a|^2 = \cos^2[\varphi(\tau)/2]$, $|b|^2 = \sin^2[\varphi(\tau)/2]$ and we obtain

$$\sigma_A = \begin{pmatrix} \cos^2(\varphi(\tau)/2) & 0 \\ 0 & \sin^2(\varphi(\tau)/2) \end{pmatrix}. \tag{19}$$

Thus, the entanglement resulting from Eq. (16) is

$$E(\tau) = -\cos^2[\varphi(\tau)/2]\log_2\{\cos^2[\varphi(\tau)/2]\} - \sin^2[\varphi(\tau)/2]\log_2\{\sin^2[\varphi(\tau)/2]\} \tag{20}$$

Comparing solutions of Eqs. (11), (12) and (20) (see Fig. 1) it is possible to draw a conclusion, that dynamics of entanglement is directly correlated with dynamics of the intensity of MQ coherences of the second order.

Let us consider dynamics of the entanglement in terms of the concurrence, which can be determined as [25, 26]

$$C(\psi) = \left|\sum_i \alpha_i^2\right|, \tag{21}$$

where $\alpha_i$ are the coefficients of the expansion of some state

$$|\psi\rangle = \sum_i^4 \alpha_i\,|e_i\rangle \tag{22}$$

in the following "magic basis" of completely entangled states:

$$|e_1\rangle = \frac{1}{\sqrt{2}}(|00\rangle + |11\rangle), \qquad |e_2\rangle = \frac{1}{\sqrt{2}}i(|00\rangle - |11\rangle), \tag{23}$$

$$|e_3\rangle = \frac{1}{\sqrt{2}}i(|01\rangle + |10\rangle), \qquad |e_4\rangle = \frac{1}{\sqrt{2}}(|01\rangle - |10\rangle).$$

Then the entanglement of $|\psi\rangle$ is [25, 26]

$$E(\psi) = H\left[\tfrac{1}{2}(1 + \sqrt{1 - C^2})\right], \tag{24}$$

where $H$ is the binary entropy function $H(x) = -x\log_2(x) - (1-x)\log_2(1-x)$.

The Eq. (13) can be rewritten in the form (22) as

$$|\Psi\rangle = \frac{1}{2}(e^{-i\varphi(\tau)/2} + ie^{-i\varphi(\tau)/2})\,\frac{1}{\sqrt{2}}(|00\rangle + |11\rangle) + \frac{1}{2}(e^{i\varphi(\tau)/2} - ie^{i\varphi(\tau)/2})\,\frac{1}{\sqrt{2}}i(|00\rangle - |11\rangle). \tag{25}$$

Thus $\alpha_1 = \frac{1}{2}(e^{-i\varphi(\tau)/2} + ie^{-i\varphi(\tau)/2})$, $\alpha_2 = \frac{1}{2}(e^{i\varphi(\tau)/2} - ie^{i\varphi(\tau)/2})$, $\alpha_3 = \alpha_4 = 0$ and

$$C = |\alpha_1^2 + \alpha_2^2| = |\sin\varphi(\tau)|. \tag{26}$$

Comparing Eqs. (12) and (26) it is possible to conclude that in the two-spin system the concurrence is connected with the intensity of MQ coherences of the second order by the relation:

$$C^2(\tau) = J_2(\tau). \quad (27)$$

Note that the entanglement determined by Eqs. (20) and (24) is the same.

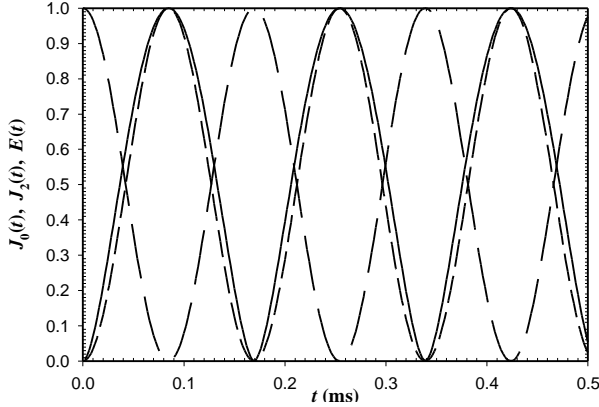

FIG. 1. Time dependence of the intensity of MQ coherences of the zeroth- order $J_0(t)$ (long-dash line), of the second order $J_2(t)$ (the square of the concurrence $C^2(t)$) (short-dash line), of entanglement $E(t)$ (solid line) for two spins coupled by DDI with the dipolar coupling constant $D_{12}=2\pi\,2950$ s$^{-1}$.

## IV. THREE-SPIN CASE

It is known [11] that the operator $e^{i\pi I_z}$ is an integral of motion. Hence, the Hamiltonian ($2^N \times 2^N$) matrix is reduced to two $2^{N-1} \times 2^{N-1}$ submatrices which correspond to odd and even numbers of the spins which are directed opposite to the applied magnetic field $\vec{H}_0$. If the number of spins is odd then both blocks give the same profiles of MQ coherences. One should solve the problem using one block of the density matrix and double the intensities of MQ coherences [11]. If the number of spins is even then blocks give the different profiles of MQ coherences, but they can be calculated independently one from other [11].

For three-spin systems, there are eight possible product states: $|000\rangle$, $|001\rangle$, $|010\rangle$, $|011\rangle$, $|100\rangle$, $|101\rangle$, $|110\rangle$, $|111\rangle$. The density matrix is reduced to two blocks in basis $|000\rangle$, $|011\rangle$, $|101\rangle$, $|110\rangle$ (for even number of individual spin states $|1\rangle$) and $|001\rangle$, $|010\rangle$, $|100\rangle$, $|111\rangle$ (for odd number of such spins) correspondingly.

The top block of the initial density matrix according to Eq. (5) is

$$\rho(0) = \begin{pmatrix} \frac{3}{2} & 0 & 0 & 0 \\ 0 & -\frac{1}{2} & 0 & 0 \\ 0 & 0 & -\frac{1}{2} & 0 \\ 0 & 0 & 0 & -\frac{1}{2} \end{pmatrix}. \quad (28)$$

The time development of the density matrix can be expressed analytically [22]:

$$\rho(\tau) = \rho_0(\tau) + \rho_{\pm 2}(\tau) = \begin{pmatrix} \frac{3}{2}+A & iR_{23} & iR_{13}B & iR_{12}B \\ -iR_{23}B & -\frac{1}{2}-R_{23}^2 A & -R_{13}R_{23}A & -R_{12}R_{23}A \\ -iR_{13}B & -R_{13}R_{23}A & -\frac{1}{2}-R_{13}^2 A & -R_{12}R_{13}A \\ -iR_{12}B & -R_{12}R_{23}A & -R_{12}R_{13}A & -\frac{1}{2}-R_{12}^2 A \end{pmatrix}, \quad (29)$$

where $A=[\cos(D_{\text{eff}}\,\tau) - 1]$, $B= \sin(D_{\text{eff}}\,\tau)$ and $R_{12}=D_{12}/D_{\text{eff}}$, $R_{13}=D_{13}/D_{\text{eff}}$, $R_{23}=D_{23}/D_{\text{eff}}$. Here, $D_{\text{eff}}=(D^2_{12} + D^2_{13} + D^2_{23})^{1/2}$.

Consider a situation where all three coupling constants are equal (the triangle ring) $D= D_{12}= D_{13}= D_{23}$. In this case, $D_{\text{eff}}=\sqrt{3}\,D$, $R=R_{12}= R_{13}= R_{23}=\dfrac{1}{\sqrt{3}}$ and the density matrix simplifies to

$$\rho(\tau)=\begin{pmatrix} \tfrac{3}{2}+A & i\tfrac{1}{\sqrt{3}}B & i\tfrac{1}{\sqrt{3}}B & i\tfrac{1}{\sqrt{3}}B \\ -i\tfrac{1}{\sqrt{3}}B & -\tfrac{1}{2}-\tfrac{1}{3}A & -\tfrac{1}{3}A & -\tfrac{1}{3}A \\ -i\tfrac{1}{\sqrt{3}}B & -\tfrac{1}{3}A & -\tfrac{1}{2}-\tfrac{1}{3}A & -\tfrac{1}{3}A \\ -i\tfrac{1}{\sqrt{3}}B & -\tfrac{1}{3}A & -\tfrac{1}{3}A & -\tfrac{1}{2}-\tfrac{1}{3}A \end{pmatrix}, \quad (30)$$

where $A=[\cos\varphi(\tau) - 1]$, $B= \sin\varphi(\tau)$ and $\varphi(\tau)=\sqrt{3}\,D\tau$.

Then the intensity of the normalized MQ coherences of the second order (for both blocks) can be calculated [22] from Eqs. (7) and (30) as

$$J_2(\tau)=\frac{2}{3}\sin^2\varphi(\tau). \quad (31)$$

As well as in the case of a two-spin system, the density matrix $\rho(\tau)$ of Eq. (30) can be expressed through the vector of the state

$$|\Psi\rangle= e^{i\tfrac{\pi}{4}}\cos[\varphi(\tau)/2]\,|000\rangle + \frac{1}{\sqrt{3}}e^{-i\tfrac{\pi}{4}}\sin[\varphi(\tau)/2]\,|011\rangle +\frac{1}{\sqrt{3}}e^{-i\tfrac{\pi}{4}}\sin[\varphi(\tau)/2]\,|101\rangle +$$

$$+\frac{1}{\sqrt{3}}e^{-i\tfrac{\pi}{4}}\sin[\varphi(\tau)/2]\,|110\rangle. \quad (32)$$

The projector is

$$\sigma(\tau)=|\Psi\rangle\langle\Psi|=\frac{1}{2}\begin{pmatrix} 2+A & i\tfrac{1}{\sqrt{3}}B & i\tfrac{1}{\sqrt{3}}B & i\tfrac{1}{\sqrt{3}}B \\ -i\tfrac{1}{\sqrt{3}}B & -\tfrac{1}{3}A & -\tfrac{1}{3}A & -\tfrac{1}{3}A \\ -i\tfrac{1}{\sqrt{3}}B & -\tfrac{1}{3}A & -\tfrac{1}{3}A & -\tfrac{1}{3}A \\ -i\tfrac{1}{\sqrt{3}}B & -\tfrac{1}{3}A & -\tfrac{1}{3}A & -\tfrac{1}{3}A \end{pmatrix}. \quad (33)$$

Comparing (30) and (33), we have

$$\rho(\tau)=2\sigma(\tau) - \frac{1}{2}E, \quad (34)$$

where $E$ is the unit matrix.

To find connection between MQ dynamics and dynamics of the entanglement for a three-spin system we shall analyze the entanglement properties of the pure state (32) through the state

$$|\Phi\rangle = a\,|000\rangle + b\,|011\rangle + c\,|101\rangle + d\,|110\rangle, \quad (35)$$

where $|a|^2 + |b|^2 + |c|^2 + |d|^2 = 1$.

The density matrix of three particles $A,B,C$ is $\sigma_{ABC}= |\Phi(ABC)\rangle\langle\Phi(ABC)|$. The two-particle and one-particle reduced density matrices are given by

$$\sigma_{BC}=\text{Tr}_A\sigma_{ABC},$$
$$\sigma_A=\text{Tr}_{BC}\sigma_{ABC}. \quad (36)$$

All other reduced density matrices are obtained analogically.

For the state (35) we have

$$\sigma_{BC} = \begin{pmatrix} |a|^2 & 0 & 0 & ab^* \\ 0 & |c|^2 & cd^* & 0 \\ 0 & dc^* & |d|^2 & 0 \\ ba^* & 0 & 0 & |b|^2 \end{pmatrix}, \quad \sigma_A = \begin{pmatrix} |a|^2 + |b|^2 & 0 \\ 0 & |c|^2 + |d|^2 \end{pmatrix}. \quad (37)$$

The matrices $\sigma_{AC}$, $\sigma_B$, and $\sigma_{AB}$, $\sigma_C$ are obtained from (37) by the changes $c \Leftrightarrow b$ and $c \Leftrightarrow d$ performed one after the other.

For the analysis of the entanglement we use the approach advanced in Ref. [27]. We shall introduce the "spin-flipped" density matrix $\tilde{\sigma}_{BC} = (\sigma_y \otimes \sigma_y) \sigma^*_{BC} (\sigma_y \otimes \sigma_y)$, where $\sigma^*_{BC}$ is the complex-conjugate matrix of $\sigma_{BC}$ and $\sigma_y$ is the standard y- component Pauli matrix. As both $\sigma_{BC}$ and $\tilde{\sigma}_{BC}$ are positive operators, it follows that the product $\sigma_{BC} \tilde{\sigma}_{BC}$, though non-Hermitian, also has only real and non-negative egenvalues [27].
Then

$$\sigma_{BC} \tilde{\sigma}_{BC} = \begin{pmatrix} 2|a|^2|b|^2 & 0 & 0 & 2|a|^2 ab^* \\ 0 & 2|c|^2|d|^2 & 2|c|^2 cd^* & 0 \\ 0 & 2|d|^2 dc^* & 2|c|^2|d|^2 & 0 \\ 2|b|^2 ba^* & 0 & 0 & 2|a|^2|b|^2 \end{pmatrix} \quad (38)$$

and the two nonzero eigenvalues of this matrix product are

$$\lambda_1^{BC} = 4|a|^2|b|^2, \quad \lambda_2^{BC} = 4|c|^2|d|^2. \quad (39)$$

Having executed the above-mentioned replacement, we shall obtain

$$\lambda_1^{AC} = 4|a|^2|c|^2, \quad \lambda_2^{AC} = 4|b|^2|d|^2,$$
$$\lambda_1^{AB} = 4|a|^2|d|^2, \quad \lambda_2^{AB} = 4|b|^2|c|^2.$$

The concurrence between two particles is defined in this case as $C = |\sqrt{\lambda_1} - \sqrt{\lambda_2}|$ [26] and the square of it is

$$C^2_{BC} = 4(|a||b| - |c||d|)^2, \quad C^2_{AC} = 4(|a||c| - |b||d|)^2, \quad C^2_{AB} = 4(|a||d| - |b||c|)^2. \quad (40)$$

The concurrence between particle $A$ and the pair $BC$ can be calculated [27] as
$$C^2_{A(BC)} = \mathrm{Tr}\sigma_{AB}\tilde{\sigma}_{AB} + \mathrm{Tr}\sigma_{AC}\tilde{\sigma}_{AC} = 4\det\sigma_A, \quad (41)$$
and is similar in remaining cases.
We obtain

$$C^2_{A(BC)} = 4(|a|^2|c|^2 + |a|^2|d|^2 + |b|^2|c|^2 + |b|^2|d|^2),$$
$$C^2_{B(AC)} = 4(|a|^2|b|^2 + |a|^2|d|^2 + |b|^2|c|^2 + |c|^2|d|^2),$$
$$C^2_{C(AB)} = 4(|a|^2|b|^2 + |a|^2|c|^2 + |b|^2|d|^2 + |c|^2|d|^2). \quad (42)$$

One can easily verify that
$$C^2_{A(BC)} - C^2_{AB} - C^2_{AC} = C^2_{B(AC)} - C^2_{AB} - C^2_{BC} = C^2_{C(AB)} - C^2_{AC} - C^2_{BC} =$$
$$= 16|a||b||c||d| = \tau_{ABC}, \quad (43)$$

where the three-tangle [27] $\tau_{ABC}$ is the measure of three-particle entanglement. Apparently, the pure state for the second block for an odd number of spins

$$|\Phi\rangle = d\,|001\rangle + c\,|010\rangle + b\,|100\rangle + a\,|111\rangle \qquad (44)$$

has analogous measures of the entanglement.

It is interesting to compare measures of entanglement of the state given by Eq. (35) or Eq. (44) with that for other classes of pure states. One can see, that these expressions generalize the states known as *GHZ* and *W*. Depending on a value of coefficients they have measures of entanglement relevant either for the *GHZ* or *W* state.

For example, under the condition that $|a|^2=|b|^2=|c|^2=|d|^2=\frac{1}{4}$, all pair concurrences $C_{BC}^2$, $C_{AC}^2$, $C_{AB}^2$ are equal to zero and $C_{A\,(BC)}^2 = C_{B\,(AC)}^2 = C_{C\,(AB)}^2 = \tau_{ABC}=1$ similar to the *GHZ* state [27]. In this case we have the maximal three-spin entanglement. However the spins of each pair are classically correlated but not entangled. It also implies that when one of the spins is traced out then the remaining two are unentangled.

If only one of the coefficients $a$, $b$, $c$, $d$ is zero we have a state analogous to the *W* state. This is obvious when $a=0$ but it is non-trivial in remaining cases. In all these cases the three-tangle $\tau_{ABC}$ is equal to zero; the condition $C_{A\,(BC)}^2 = C_{AB}^2 + C_{AC}^2$ is also satisfied (analogically for *B–AC*, *C–AB*). It means that this case corresponds to the *W* state [27].

In general case the state (35) or (44) has all characteristics of the entanglement which are not equal to zero simultaneously. This property distinguishes these states from other known pure states.

Returning to the vector state of Eq. (32) we have time-periodical coefficients. Then the miscellaneous states will be realized including the completely separable state when $\sin[\varphi(\tau)/2]=0$.

Three-spin states of Eqs. (35), (44) can be generalized for an arbitrary number of spins. For example, the well-known Bell basis for the two-spin systems $|\psi^\pm\rangle = \frac{1}{\sqrt{2}}(|00\rangle \pm |11\rangle)$, $|\phi^\pm\rangle = \frac{1}{\sqrt{2}}(|01\rangle \pm |10\rangle)$ is also the special case of these states with an even and odd number of $|1\rangle$ single-spin states accordingly. The case when the total number of spins is even is different from the one with odd number of spins. Two states transfer one to the other with flip of all spins when their number is odd. The analogous procedure does not change these states at even number of spins.

For our problem we have from Eqs. (32), (39)

$$\lambda_1 = \lambda_1^{BC} = \lambda_1^{AC} = \lambda_1^{AB} = \frac{1}{3}\sin^2\varphi(\tau),$$

$$\lambda_2 = \lambda_2^{BC} = \lambda_2^{AC} = \lambda_2^{AB} = \frac{4}{9}\sin^4[\varphi(\tau)/2] \qquad (45)$$

and from Eqs. (40-43) we obtain

$$C^2_{BC} = C^2_{AC} = C^2_{AB} = (\sqrt{\lambda_1} - \sqrt{\lambda_2})^2 = (\frac{1}{\sqrt{3}}|\sin\varphi(\tau)| - \frac{2}{3}\sin^2[\varphi(\tau)/2])^2,$$

$$C^2_{A(BC)} = C^2_{B(AC)} = C^2_{C(AB)} = 2(\lambda_1+\lambda_2) = 2(\frac{1}{3}\sin^2\varphi(\tau) + \frac{4}{9}\sin^4[\varphi(\tau)/2]),$$

$$\tau_{ABC} = 4\sqrt{\lambda_1\lambda_2} = \frac{8}{3\sqrt{3}}|\sin\varphi(\tau)|\sin^2[\varphi(\tau)/2]. \tag{46}$$

If we have the condition $b=c=d\neq a$ as in our case, then it is enough to know only one eigenvalue $\lambda_1$ for the full solution of the problem of the entanglement. Another eigenvalue $\lambda_2$ can be obtained from the expression

$$\lambda_1 = 2\sqrt{\lambda_2} - 3\lambda_2. \tag{47}$$

Comparing Eq. (31) with Eq. (45) we conclude, that the intensity of MQ coherences of the second order coincides with the eigenvalue $\lambda_1$ (for both blocs with the double $\lambda_1$)

$$J_2(\tau) = 2\lambda_1. \tag{48}$$

The same result was obtained before for a two-spin system in the previous section. Then we had one block only (another block is zero) and $J_2(\tau) = C^2(\tau) = \lambda$ (the matrix $\sigma\tilde\sigma$ has one nonzero eigenvalue). Thus, it is possible to conclude that the intensity of MQ coherences of the second order is the generalized measure of entanglement. Dynamics of entanglement can be investigated from the time evolution of MQ coherences.

Fig. 2 shows the evolution of concurrences and the three-tangle for the three-spin ring. The evolution of the corresponding entanglements (Eq. (24)) has the similar forms and they are represented on Fig. 3. All these curves can be obtained from the time evolution of the intensity of MQ coherences of the second order (Fig. 4) according to Eqs. (48), (47), (46), (24).

From Figs. 2,3 we can see that the completely separable state (all measures of the entanglement equal to zero) takes place on times $\tau = 0.1957$ ms, $0.3914$ ms, … when $\sin[\varphi(\tau)/2] = 0$.

For systems with larger number of spins, the situation is more complicated. MQ dynamics of such systems includes the MQ coherences of higher orders. It is possible to assume that dynamics of entanglement in this case is determined by the intensities of MQ coherences of orders 2, 6, 10,…, $\leq N$ which can be described by the imaginary part [11] of the density matrix.

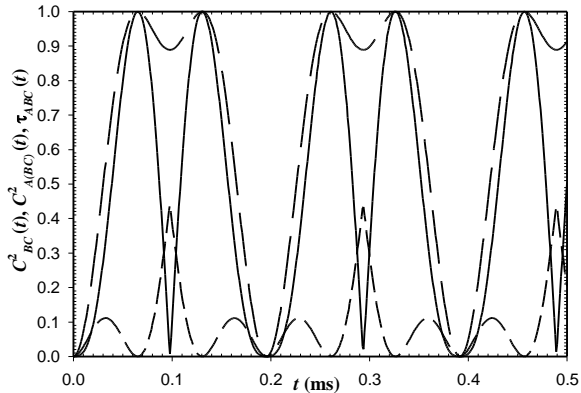

FIG. 2. Time dependence of the square of the pairwise concurrence $C^2_{BC}(t)$ (short-dash line), the square of the concurrence $C^2_{A(BC)}(t)$ (long-dash line), the three-tangle $\tau_{ABC}(t)$ (solid line) for three spins coupled by DDI with the dipolar coupling constant $D_{12} = D_{23} = D_{13} = 2\pi\, 2950$ s$^{-1}$.

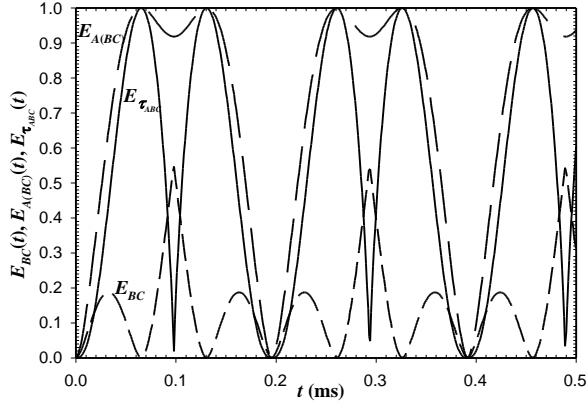

FIG. 3. Time dependence of the pairwise entanglement $E_{BC}(t)$ (short-dash line), the entanglement $E_{A(BC)}(t)$ (long-dash line), the three-tangle entanglement $E\tau_{ABC}(t)$ (solid line) for three spins coupled by DDI with the dipolar coupling constant $D_{12}= D_{23}= D_{13}=2\pi\,2950$ s$^{-1}$.

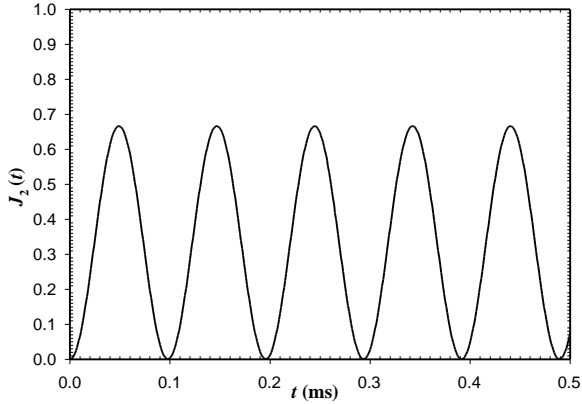

FIG. 4. Time dependence of the intensity of MQ coherence of the second order $J_2(t)$ for three spins coupled by DDI with the dipolar coupling constant $D_{12}= D_{23}= D_{13}=2\pi\,2950$ s$^{-1}$.

## ACKNOWLEDGMENTS

The author thanks Edward B. Fel'dman for helpful discussions. The Russian Foundation for Basic Research under Grant 01-03-33273 and the Moscow Region Administration under Grant 01-01-97002 support the work.